# Avoid Internal Loops in Steady State Flux Space Sampling

Lu Xie

**ABSTRACT:** As a widely used method in metabolic network studies, Monte-Carlo sampling in the steady state flux space is known for its flexibility and convenience of carrying out different purposes, simply by alternating constraints or objective functions, or appending post processes. Recently the concept of a non-linear constraint based on the second thermodynamic law, known as "Loop Law", is challenging current sampling algorithms which will inevitably give rise to the internal loops. A generalized method is proposed here to eradicate the probability of the appearance of internal loops during sampling process. Based on Artificial Centered Hit and Run (ACHR) method, each step of the new sampling process will avoid entering "loop-forming" subspaces. This method has been applied on the metabolic network of *Helicobacter pylori* with three different objective functions: uniform sampling, optimizing biomass synthesis, optimizing biomass synthesis efficiency over resources ingested. Comparison between results from the new method and conventional ACHR method shows effective elimination of loop fluxes without affecting non-loop fluxes.

**INTRODUCTION**

As a constraint-based modeling approach, Flux Balance Analysis (FBA) [1] has been widely used [2] to investigate the structure-function relations of metabolic networks. In FBA, the flux vectors satisfying the steady-state requirements constitute a solution space, on which additional constraints could be imposed. The imposition of linear constraints, such as physicochemical reversibility and biological capacity, shape the solution space into a convex cone [3], whereas linear programming can be used to optimize linear objective functions in the constrained solution space, for instances, to predict the optimal growth rates [4], to measure ranges of achievable flux values [5], and to minimize the stationary metabolic fluxes [6]. However, the imposition of non-linear constraints or the optimization of non-linear objectives in FBA requires more than linear approaches. One complementary is sampling-based methods. They not only have the advantages in the imposition of non-linear constraints or non-linear optimization objectives, but also provide ensembles of fluxes which reflect possible living states of the organism, and thus also have been employed widely in metabolic network analysis [7, 13-14, 26-27].

One of the non-linear constraints, termed as the "Loop Law" [9-10], has drawn attention from the users and developers of sampling methods. For those metabolic networks which are able to form internal loops (Fig. 1A), current sampling methods cannot avoid samples containing internal loops because they satisfy the flux balance constraint and exist in the solution space. Internal loops disobey the second thermodynamic law and can only be constrained by flux limits, which boost the involved fluxes to unreasonable values and fluctuations. An additional constraint again internal loops is the imposition of "Loop Law", which separates the solution space into "loop-free" and "loop-forming" subspaces (Fig. 1B), identified by the direction patterns of the reversible reactions in the loop. The flux vectors in the "loop-free" subspaces could satisfy the constraint of "Loop Law" [10-11]. One heuristic attempt of introducing "Loop Law" to sampling in the solution space was shrinking the maximum and minimum boundaries of the fluxes in the loop reactions [16, 27]. This attempt can attenuate the affect of internal loops on sample ensembles, or decrease the possibility of observing a sample with internal loops, but has no guarantee of eradication; therefore it may require a post process to screen out the loop-containing sample points [16]. Besides, it also sweeps away the loop-free samples with flux exceeding the boundary.

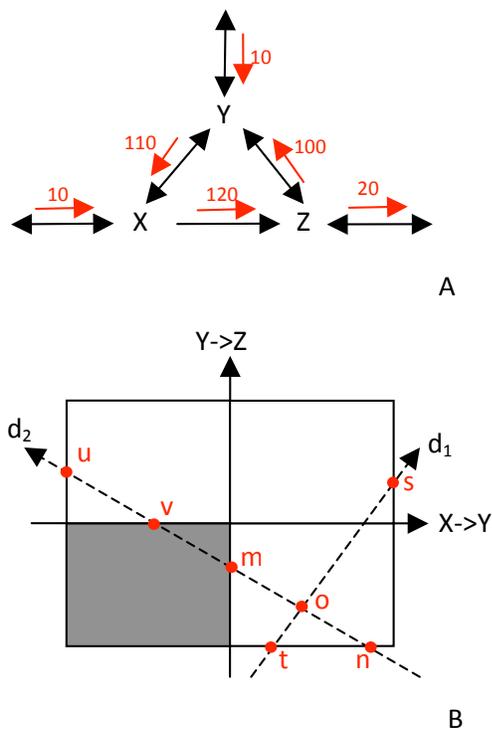

Fig. 1
A. A combination of reactions that can form an internal loop, indicated by red arrows. X, Y, Z are metabolites, the double black arrows refer to reversible reactions, and the single black arrow refers to irreversible one. Red numbers refer to fluxes between metabolites.
B. Solution space of fluxes in A, projected onto X->Y and Y->Z plane. The original solution space is bounded by the outer solid black lines. The gray box (Y->X, Z->Y) encloses a "loop-forming" subspace, while other three are "loop-free" ones. Sampling starts from red dot o, if direction $d_1$ is chosen, the next possible sample point would be randomly chosen on segment s-t; if direction $d_2$ is chosen, the next possible sample point would be randomly chosen on segment u-v and m-n. This figure is for schematic demonstration only, since the base vectors of the solution space are null vectors, not fluxes.

The basic idea of our method is to constrain the sampling process inside the "loop-free" subspaces for each step. Based on ACHR, the sampling process starts from current sample point, randomly chooses a direction, finds which portions on this direction reside in the "loop-free" subspaces, and randomly chooses a position on those portions as next sample point (Fig. 1B). Theoretically this method could avoid collecting loop-containing samples and do not affect the distribution of non-loop reactions, which will be proved later in this paper. The major tradeoff of this method is higher computing cost on finding feasible portions for each "loop-free" subspace, instead of conventional ACHR which merely finds the feasible distance for the whole solution space (Fig. 1B).

In this work, the sampling process has been performed on the steady state flux space of the genome-scale metabolic network of *H. pylori* (iIT341 GSM/GPR) [8]. The program is written and tested through MATLAB 7.10.0.499 64-bit. Four sample ensembles have been collected: 1, "Loop-Law" constraint turned off for comparison; 2, uniform sampling with "Loop-Law" turned on; 3, optimize the biomass synthesis flux; 4, optimize biomass synthetic efficiency. The motive of performing the last optimization results from the common sense that more biomass will be produced if more resource is ingested, i.e. the biomass flux will depend tightly on the upper limit of intake fluxes, which could rise a potential problem because many of these upper limits are set artificially; therefore it may be more worthwhile to figure out certain combinations of intake fluxes that maximize the biomass synthetic efficiency. Metropolis criterion and Simulated Annealing technique [15] have been employed to optimize the objective functions.

## METHODS

*Model*

As a kind of notorious human gastric pathogen, H. pylori infection has been detected in more than half cases of gastric and duodenal ulcers [20]. The genome-scale metabolic network of the H. pylori strain 26695 has been reconstructed using the revised genome annotation and new experimental data in 2005 and marked as iIT341 GSM/GPR [8], which accounts for 341 metabolic genes, 476 internal reactions, 74 external reactions, 8 demand functions, 411 internal metabolites and 74 external metabolites. The demand functions are the reactions of the type: A -->, which means that the compound A can be only produced by the network, but without further balancing the compound A. The demand functions include a Biomass function, a HMFURN function, excretion of Thiamin, Menaquinone 6, Biotin and Heme (Protoheme), and sinking reactions of ahcys(c) and amob. The external reactions and demand functions, or called exchange reactions, serve as the input and output of the network. The raw version of stoichiometric matrix, in total, has 558 reactions and 485 metabolites.

**Preprocess:** *delete dead-ends*

The original model iIT341 GSM/GPR contains dead-end metabolites and zero fluxes [8]. One way to find these reactions is to calculate the maximum and the minimum values of every flux by linear programming [5]:

$$\begin{aligned} & for\ i \in R \\ & \quad \min(-v_i)\ \&\ \min(v_i) \\ & \quad s.t. \begin{cases} S \cdot v = 0 \\ \alpha_k \leq v_k \leq \beta_k, k \in R \end{cases} \\ & end \end{aligned} \quad (1)$$

Here $S$ stands for the stoichiometic matrix, $R$ stands for the set of reactions, $α_k$ and $β_k$ denote the lower and upper bound of the $k$-th flux, respectively. The $v_i$ which is always tightly constrained to zero reflects that the i-th reaction contains at least one dead-end metabolite. Deletion of these reactions and the dead-end metabolites would reduce the dimension of the stoichiometric matrix and the size of solution space.

**Preprocess:** *find potential internal loops*

Prior to avoiding the existence of fluxes in the internal loops, the combinations of those reactions that are capable of forming loops must be located. As described above, the internal loops do not contain exchange fluxes [10]. For this character, the set of exchange reactions, noted as $R_{ex}$, were forced to zero in order to make the system closed. After that a linear programming was performed to find the maximum and minimum allowable fluxes on the remained reactions, each at a time [5]:

$$\text{for}(i \in \overline{R^x})$$
$$\min(-v_i) \ \& \ \min(v_i)$$
$$s.t. \begin{cases} S \cdot v = 0 \\ v_j = 0, j \in R^{ex} \\ \alpha_k \le v_k \le \beta_k, k \in \overline{R^{ex}} \end{cases} \quad (2)$$
end

Here the bar over a set means its complementary set. From the results, the fluxes of the reactions that are unable to form loops were constrained to zero while others were not, and these non-zero reactions may form one or several loops. To determine the least reactions for each loop, another linear programming is needed:

$$\text{for}(i \in R^{loop})$$
$$\min(\sum_{j \ne i | j \in R} |v_j'|)$$
$$s.t. \begin{cases} v_i' = v_i \\ S \cdot v' = 0 \end{cases} \quad (3)$$
end

For each loop flux $v_i$, the fluxes that remained non-zero in $v'$ are necessary for $v_i$ to form its belonging loop(s). Those reactions, each of which merely belongs to one loop, would help us to identify each of these loops.

**Preprocess:** *find "loop-free" subspaces*

The reactions set of a loop is noted as $R^{loop}$. If we arbitrarily assign one chemical potential to each metabolite in this loop, the sum of the changes in chemical potentials around a loop equals to zero:

$$\sum \Delta u_i = 0, i \in R^{loop} \quad (4)$$

With:

$$\Delta u_i = \sum S_{j,i} \cdot u_j, i \in R, j \in M \quad (5)$$

Here Δu is a vector whose element indicates the chemical potential change of each reaction, and u is a vector whose element indicates the chemical potential of each metabolite. S is the stoichiometric matrix, whose rows and columns represent the metabolites and reactions, respectively. For each reaction, the direction of its flux and the change of chemical potential should satisfy the constraint based on the laws of thermodynamics, i.e., all fluxes must follow the downward direction of the chemical potential change:

$$v_i \cdot \Delta u_i \le 0, \forall i \in R \quad (6)$$

Here *v* denotes for the flux vector and $v_i$ denotes for the $i^{th}$ flux value. When the inequalities Eq. 6 are satisfied, the net flux around a biochemical loop must be forced to zero, which reflects the essence of "loop law".

As discusses above, the formation of a loop relies on the direction combinations of the reversible reactions in this loop [9-10]. The reaction set of a specific loop, noted as $R_s^{loop}$, has $2^r$ possible direction combinations if it contains r reversible reactions, i.e. it has totally $2^r$ number of "loop-forming" and "loop-free" subspaces. To examine all the $2^r$ subspaces, one flux vector was randomly generated in each subspace as a representative. These vectors, noted as ($v^t$ | $t$ = 1…$2^r$), were lately examined by solving a set of inequalities each at a time:

$$for (\ t=1...2^r)$$
$$solve[u \mid \sum_{j \in M} u_j \cdot (S_{j,i} \cdot v_i^t) < 0; i \in R^{rev} \cap R_s^{loop}] \tag{7}$$
$$end$$

Here $R^{rev}$ denotes the set of reversible reactions and the vector u represents the set of virtual chemical potential of the metabolites. If the set of inequality is solvable under a specific direction pattern, there must be a set of feasible chemical potentials for the metabolites to satisfy the "Loop Law", i.e. this combination indicates a "loop-free" subspace. Each "loop-free" subspace is stored as two vectors: lower boundary (the $i^{th}$ element equals 0 if the $i^{th}$ reversible reaction must be positive) and upper boundary (the $i^{th}$ element equals 0 if the $i^{th}$ reversible reaction must be negative). The zero elements for irreversible reactions and all non-zero elements in the subspace boundaries come from the original biological capacity constraints.

***Preprocess:*** *cumulative modifications*

The linear programming in (1) has identified 53 dead-end reactions, which have been deleted as well as their dead-end metabolites.

Through the linear programming (2) and (3), six internal loops have been located in the original network. Comparing to other methods, such as analyzing extreme pathways and searching for type III pathways [10, 25], linear programming consumes much less computing time. Two of the six loops consist of the reaction sets {HPROa, HPROx} and {4HGLS, OCBTi, PHCHGS}, respectively. However, all of the five reactions are isolated from the other part of the network, and therefore they have been deleted along with the metabolites that only participate in these reactions.

Through solving the inequalities in (7), one of the remaining four loops, consists of {ASHERL2, METB1r, SHSL1r, SHSL2r}, yields no "loop-free" subspace. Further analysis indicates that the metabolite hcys-L is synthesized via ASHERL2r and RHCCE, but consumed via SHSL2r only, and the metabolite cyst-L appears only in METB1r and SHSL1r. As a result of that, deleting any one of METB1r, SHSL1r, or SHSL2r would cut out other one or more reactions. However, deletion of ASHERL2 would not cause any unfeasible affection of other reactions and might be considered as a better solution to dissemble the loop.

In general, the final version of the network contains 418 metabolites and 499 reactions, including 67 exchange reactions and 3 internal loops (Fig. 2):

$R_1^{Loop}$: {ACKr, HSERTA, METB1r, PTAr, SHSL1r, SHSL4r, HSK, THRD_L, THRS};

$R_2^{Loop}$: {H2CO3D, H2CO3D2, HCO3E};

$R_3^{Loop}$: {Nat3_1, PROt2r, PROt4r}.

$R_1^{loop}$ has only one "loop-free" subspace, while $R_2^{loop}$ and $R_3^{loop}$ both have four "loop-free" subspaces. The total number of "loop-free" subspace in the original solution space is 1 x 4 x 4 = 16.

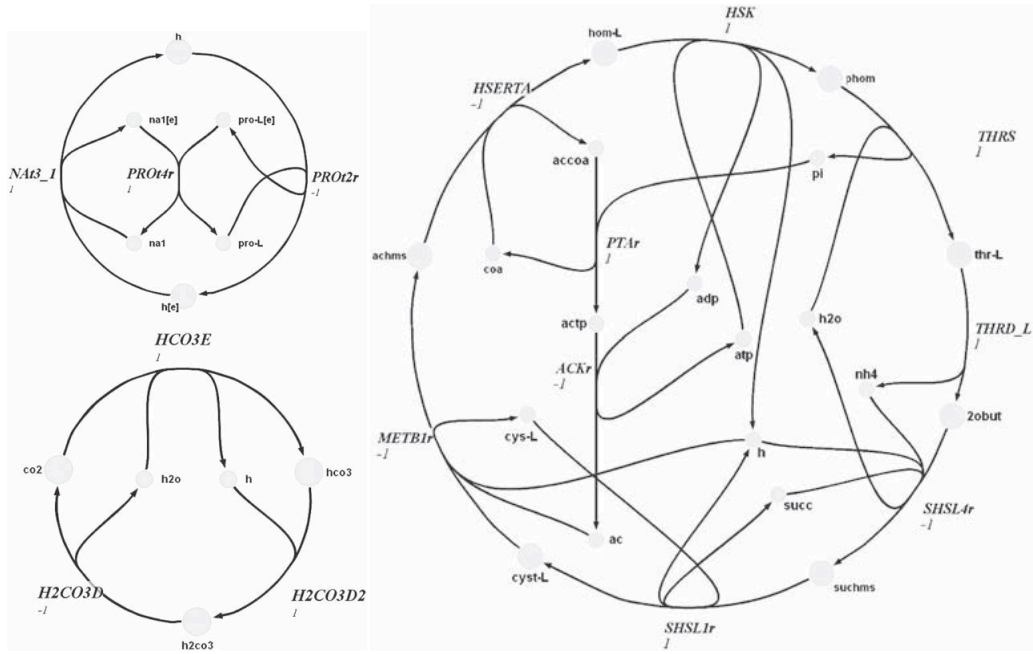

Fig. 2
Loop forming reactions in *H. pylori* metabolic network [16]

***Sampling:*** *solution space and ACHR*

Given the stoichiometric matrix *S*, with *m* rows and *n* columns, if the rank of *S* is smaller than *n*, i.e. the matrix is underdetermined, there is a null space [16, 21] in which all the flux vectors satisfy the linear equation Eq. 3. The base vector set of the null space, noted as *N*, has *n* rows and *n - rank (S)* columns and each of its columns corresponds to a base vector of the null space. The sampling was performed in the null space and its samples are represented as the coordinates of the flux vectors, noted as *c*.

$$v = N \cdot c \quad (8)$$

To impose the biological capacity constraint, each flux value has a lower limit *α* and upper limit *β*. For the reactions with known biological capacity, the values of *α* and *β* were chosen as the Ref. [11]. The value of α and β for other reactions were set to +/-10$^6$, respectively, and α ≥ 0 specifically for irreversible reactions. The vector consists of all α or all β are referred as lower or upper boundary for the solution space, *L* or *U*, respectively.

For sampling without considering of "Loop Law", Artificial Centered Hit and Run (ACHR) method [16-17] is performed in the null space constrained by *L* and *U*:

   a. Give current flux vector $v_i$;
   b. Give a direction *d* in the solution space;
   c. Find the maximum forward distance on *d* before hitting the boundaries:
      $F = \min((L^- - v^-)/d^-, (U^+ - v^+)/d^+)$
   d. Find the maximum backward distance on *d* before hitting the boundaries:
      $B = \max((L^+ - v^+)/d^+, (U^- - v^-)/d^-)$
   e. Randomly choose a position $v_{i+1}$ on *d* between *F* and *B*;
   f. Record $v_{i+1}$, update sample center *o*;
   g. Set i = i + 1, go to step a.

Here +/- superscript means the indices of positive/negative elements of vector *d*. During "warm up" process, the direction *d* in step b is randomly chosen; in normal sampling process, a sample vector *v'* is randomly chosen from current sample ensemble and set *d = v' – o*. In practice it is preferred to record sample points and update sample center for every certain steps.

To impose the constraint of "loop law", the upper and lower boundaries are replaced by the individual boundary pairs of the "loop-free" subspaces, and the forward and backward distances are calculated within the subspaces. Intuitively, the boundaries of the subspaces chop the original *d* line into segments, some of which reside in "loop-free" ones and others in "loop-forming" ones. The latter ones are ignored in calculating $v_{i+1}$ in step e.

*Sampling: objective functions*

Optimizing the biomass synthesis uses the negative value of the biomass flux as the objective function:

$$E = -v_{Biomass} \tag{9}$$

Metropolis criterion [22] was employed to find the minima of the "energy": negative biomass flux. If the new energy is lower than the old one, the new coordinate would be recorded and the next step would be started from the new coordinate. Otherwise, the acceptance rate *p* would be calculated:

$$p = \exp(\frac{-\Delta E}{T}) \tag{10}$$

Here Δ*E* is the energy change by moving every single step in the sampling and *T* is a heuristic temperature parameter. The choosing of *T* will be discussed later. The new coordinate would be accepted by the rate *p*, while by the rate (1 – *p*) it would not be recorded and the next step would start from the old coordinate.

Finding a way to define the efficiency of biomass synthesis is an open question. One possible choice would be divide the biomass production by the sum of Gibbs free energy of all ingested resources, which seems to be impractical. Here the biomass flux was divided by the total amount of reducible electrons in the substrate influxes:

$$E^u = -\frac{v_{biomass}}{\sum_{i \in R^{red}} v_i \cdot e_i} \tag{11}$$

Here $R^{red}$ denotes for the set of exchange reactions whose substrates contain reducible electrons, and $e_i$ is the number of reducible electrons of its corresponding substrate. For example, a glucose molecule has 24 reducible electrons compared to its final oxidized products: water and carbon dioxide.

*Sampling:* Simulated Annealing

For evaluation of choosing a proper temperature parameter, a random uniform sampling would be performed for certain steps without any rejections, and the energy change for each step would be recorded. The initial temperature $T_0$ was chosen as follows [23-24]:

$$T_0 = \frac{\Delta E^+}{\ln \frac{n_2}{n_2 p_0 - n_1(1 - p_0)}} \tag{12}$$

Where:

$$\Delta E^+ = \frac{1}{n_2} \sum_i \frac{\Delta E_i + |\Delta E_i|}{2} \tag{13}$$

Here $p_0$ is the initial acceptance, $n_1$ and $n_2$ are the numbers of steps in which energy alternations are negative and positive, respectively. Therefore, $\Delta E^+$ represents the average energy change for the steps which cause energy rise.

In the objective function oriented sampling process the temperature parameter would be declined to a portion of its previous value after certain steps.

**RESULT**

For each of the four sampling conditions, 10 ensembles were collected with different start points. Each ensemble consists of 100,000 sample vectors, first 5,000 of which are recorded during "warm-up" process. Each sample vector is recorded for every 100 steps. For the objective function oriented sampling processes, the initial acceptance rate was set to 0.99 and the temperature declines by 0.95 for every 1,000 samples.

First we categorized the sample vectors from the uniform sampling into "loop-free" subspaces. The 1 million sample vectors from sampling process without "Loop Law" constraint hit no "loop-free" subspace. The sample vectors under "Loop Law" constraints showed a biased distribution over the 16 "loop-free" subspaces (Fig.3), while no sample here is found to be outside of "loop-free" subspaces.

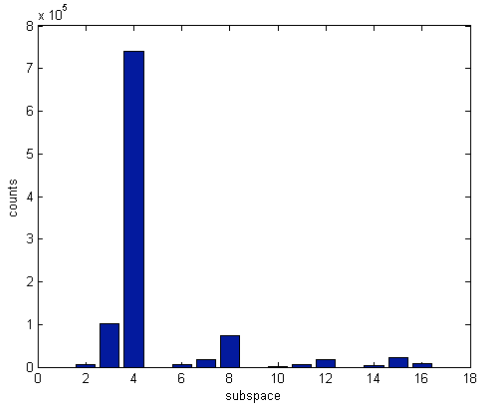

Fig. 3
The counts of samples in each "loop-free" subspace. The samples come from uniform sampling under "Loop Law" constraint.

To compare the flux distribution before and after imposing "Loop Law", the last 1,000 sample vectors are truncated from 10 ensembles of each condition and merged as one representing ensemble for histogram studies. Biomass flux was chosen to demonstrate that the imposing of "Loop Law" does not affect the distribution of non-loop reaction (Fig. 4, first row), while the flux distributions of the loop reactions changed significantly (Fig. 4, others).

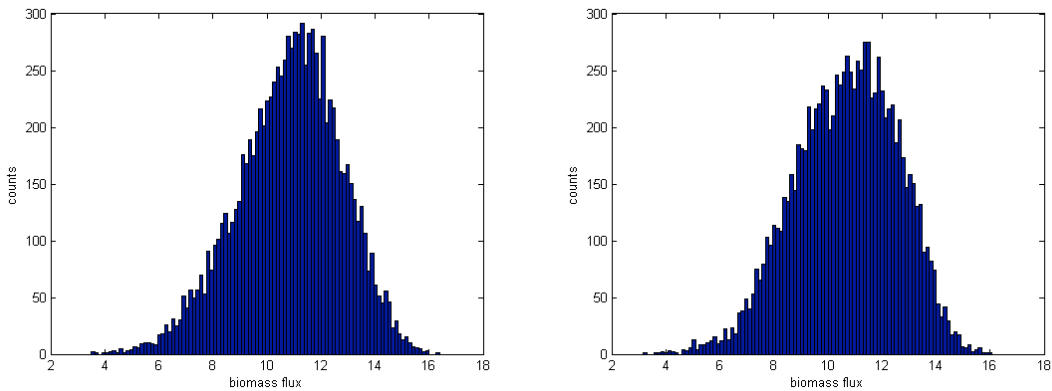

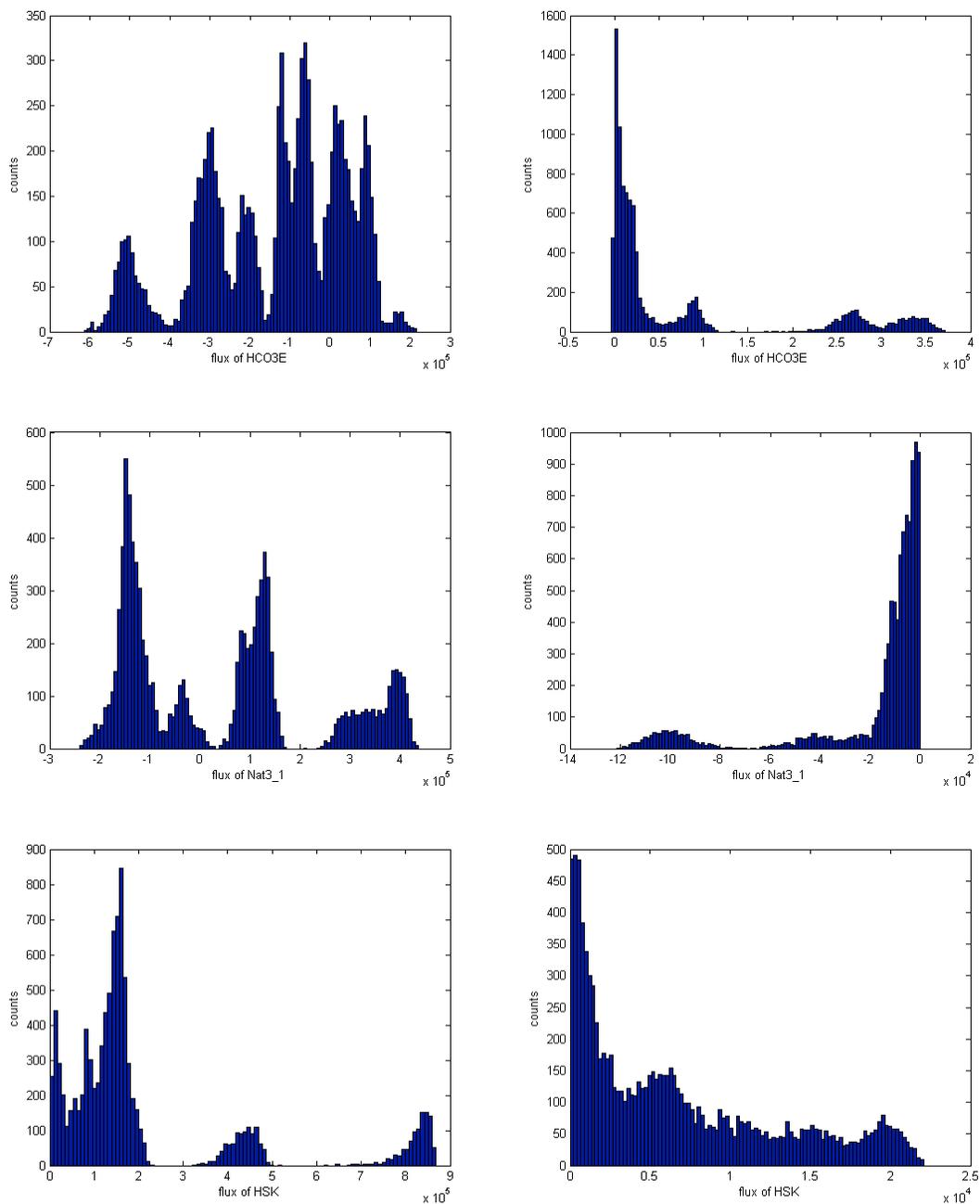

Fig. 4
Comparison of flux distributions with 100 bins
Left column: without "Loop Law"; Right column: with "Loop Law"
Row 1~4: biomass, HCO3E, Nat3_1, HSK; the latter three are loop reactions.

From the 10 ensembles of biomass optimization, the biomass flux from the highest biomass production ensemble is plotted in Fig. 5A. As expected, it does not achieve the same synthetic efficiency as the ensemble of biomass synthetic efficiency optimization (Fig 5B, C).

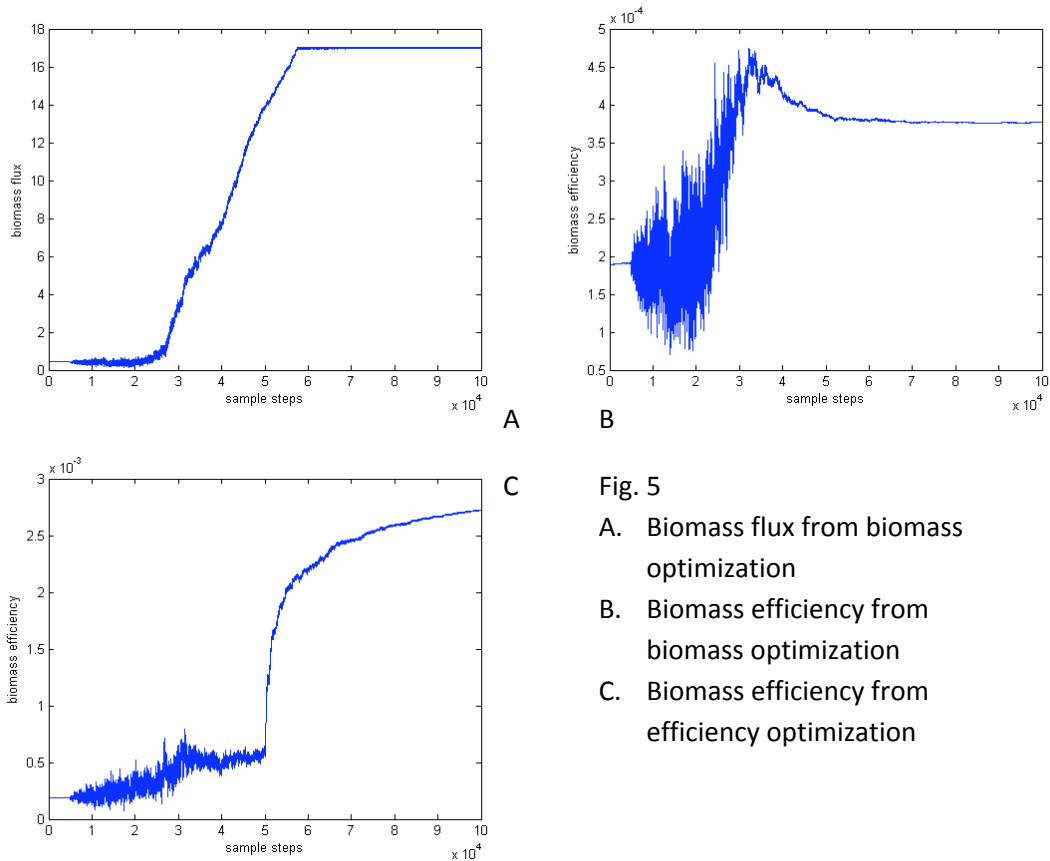

Fig. 5
A. Biomass flux from biomass optimization
B. Biomass efficiency from biomass optimization
C. Biomass efficiency from efficiency optimization

**DISCUSSION**

The method presented above successfully avoided the chance of confronting sample with internal loops while remained the distribution of other reactions unaffected. Theoretically it can be applied on any type of internal loops without too much loop structural specific intervention. It also demonstrates a more economical usage of random numbers because no sample points are thrown away. Comparing to previous work [16], the fluxes in loop reactions has similar close-to-zero distribution, but can reach higher values in our results. The previous work [16] has also reported unfavorable direction combinations of reversible loop reactions, which explains the biased distribution of subspace counts. The flux distribution of HSK, an irreversible loop reaction, has retracted to lower region after imposing "Loop Law" as the consequence of internal loop flux removal.

There still remains a large area for future investigations. The sampling results show poor consistency among ensembles with different start points, which may result from insufficient sampling steps. The parameter tweaking in simulated annealing has not been completed due to short time budget, since sharp temperature drop will confine the sampling process by a narrow space. We cannot assert our method can find the global minimum while linear programming in MATLAB will return false results if boundaries differ from 10 to $10^6$. Aside from methodological part, it is also worthwhile to study the results from biological side of view. One question would be how the removal of internal loops affects the correlation between loop and non-loop reactions; another focus would be the biased distribution of "loop-free" subspaces.

The trade-off of our method roots in extra computing costs on finding distances inside "loop-free" subspaces. Current algorithm finds all boundary hits of the line on chosen direction, for example, the direction $d_1$ in Fig. 1B will hit the boundaries of the left two boxes, although those hit points reside outside the two boxes; therefore it requires an additional proof procedure on each hit point. Running on a Dual Quad-Core Xeon E5520 2.27GHz machine, the uniform sampling process without "Loop Law" takes about 1.5 hours, while the one with "Loop Law" needs about 4.8 hours. A network with more "loop free" subspaces would consume longer time. A better distance finding algorithm can improve the computational efficiency.